\documentclass{iaus}
\usepackage{graphicx}

\title[Massive Black holes] 
{Massive Black Holes: formation and evolution}

\author[Rees \& Volonteri]   
{Martin J. Rees$^1$,
 Marta Volonteri$^2$}

\affiliation{$^1$Institute of Astronomy, University of Cambridge, \break
Madingley Road, Cambridge CB3 0HA, UK \break email: mjr@ast.cam.ac.uk\\[\affilskip]
$^2$Department of Astronomy, University of Michigan, \break 500 Church Street, Ann Arbor, MI 48109, USA \break email: martav@umich.edu}

\pubyear{2006}
\volume{238}  
\pagerange{001--999}
\date{??? and in revised form ???}
\jname{Black Holes: from Stars to Galaxies -- across the Range of Masses}
\editors{V. Karas \& G. Matt, eds.}
\begin{document}

\maketitle

\begin{abstract}
Supermassive black holes are nowadays believed to reside in most local galaxies. Observations have revealed us vast information on the population of local and distant black holes, but the detailed physical properties of these dark massive objects are still to be proven. Accretion of gas and black hole mergers play a fundamental role in determining the two parameters defining a black hole: mass and spin.  We briefly review here the basic properties of the population of supermassive black holes, focusing on the still mysterious formation of the first massive black holes, and their evolution from early times to now.   
\end{abstract}

\firstsection 
\section{Introduction}
Black holes (BHs) are the theme of this conference, spanning the full range of masses that we encounter in the astrophysical context: from tiny holes not much more massive than our sun, to monsters weighting by themselves almost as much as a dwarf galaxy. Notwithstanding the several orders of magnitude difference between the smallest and the largest BH known, we believe that all of them can be described by only three parameters: mass, spin and charge (which is not astrophysically relevant, anyway).  These in principle simple systems, however, are implicated in the power output from active galactic nuclei (AGNs), and in the production of relativistic jets that energise strong radio sources. Simple inner engines give rise to very messy and complicated energetic phenomena. 

By the early 1970s the astrophysically-relevant theoretical properties of black holes -- the Kerr metric, the 'no hair' theorems, and so forth -- were well established. By now, the evidence points insistently towards the existence of dark objects, with deep potential wells and 'horizons' through which matter can pass into invisibility. 

We expect the same mathematical and physical properties to describe both the black holes of a few solar masses and the supermassive black holes (SMBHs) in the mass range of million to billion solar masses, but while the formation path of stellar mass BHs as remnants of supernovae is by now commonly accepted, the formation and evolution of the supermassive variety  is far less understood. 

It is worth mentioning that in between the observed populations of stellar mass BHs (up to a few tens solar masses) and supermassive BHs (the smallest, detected in the dwarf Seyfert~1 galaxy POX 52, is $ \sim 10^5\,M_\odot$, Barth et al. 2004), there might exist an intermediate league, in the range of hundreds or thousands solar masses, bridging the gap. 

In the following we will focus on the formation and evolution supermassive black holes, but we remark once again that stellar-mass and supermassive holes differ for their histories, but not for their physical properties.

\section{Probing the properties of Massive Dark Objects:  black holes? spinning?}
Observers cannot yet definitively confirm the form of the metric in the strong-gravity region, in order to prove that BHs are indeed described by the Kerr metric. The flow patterns close to the hole offer, in principle, a probe of the metric. Until X-ray interferometry is developed, actually `imaging' the inner discs is beyond the capabilities of current instruments.  However, the motion of material in a relativistic potential is characterized by large frequency shifts: substantial gravitational redshifts would be expected, as well as large doppler shifts. These large frequency-shifts can be revealed spectroscopically. The dominant spectral feature in the X--ray spectrum  (2-10 keV)  is the Fe K$\alpha$ line at 6.4 keV, which is typically observed with broad asymmetric profile indicative of a relativistic disc (Fabian et al. 1989; Laor 1991).  The iron line can in principle constrain also the value of black hole spins.  Let us define the dimentionless black hole spin, $\hat a \equiv J_h/J_{max}=c \, J_h/G \,{\cal M}_{\rm BH}^2$, where 
$J_h$ is the angular momentum of the black hole.  The value of $\hat a$  affects the location of the inner radius of the accretion disc (corresponding to the innermost stable circular orbit in the standard picture), which in turn has a large impact on the shape of the line profile,  because when the hole is rapidly rotating,  the emission is concentrated closer in,  and the line displays larger shifts. There is some evidence that this must be the case  in some local Seyfert galaxies (Miniutti, Fabian \& Miller 2004, Streblyanska et al. 2005).  The assumption that the inner disc radius corresponds to the ISCO is not a trivial one, however, especially for thick discs (e.g. Krolik 1999, but see Afshordi \& Paczynski 2003).  

The spin of a hole affects the efficiency of 'classical' accretion processes themselves; the value of $\hat a$ in a Kerr BH also determines how much energy is in principle extractable from the hole itself.  Assuming that relativistic jets are powered by rotating black holes through the Blandford-Znajek
mechanism,  the orientation of the spin axis may be important in relation to jet production. 
Spin-up is a natural consequence of prolonged disc-mode accretion: any hole that has (for instance) doubled its mass by capturing material with constant angular momentum axis would end up with spinning rapidly,  close to the maximum allowed value (Thorne 1974). A hole that is the outcome of a merger between two of comparable mass would also, generically, have a substantial spin. On the other hand, a BH which had gained its mass from capturing many low-mass objects (holes, or even stars) in randomly-oriented orbits, would keep a small spin (Figure 1; see Moderski, Sikora \& Lasota 1998; Volonteri et al. 2005; Volonteri, Sikora \& Lasota 2007).
\begin{figure}
\includegraphics[width=3.5in]{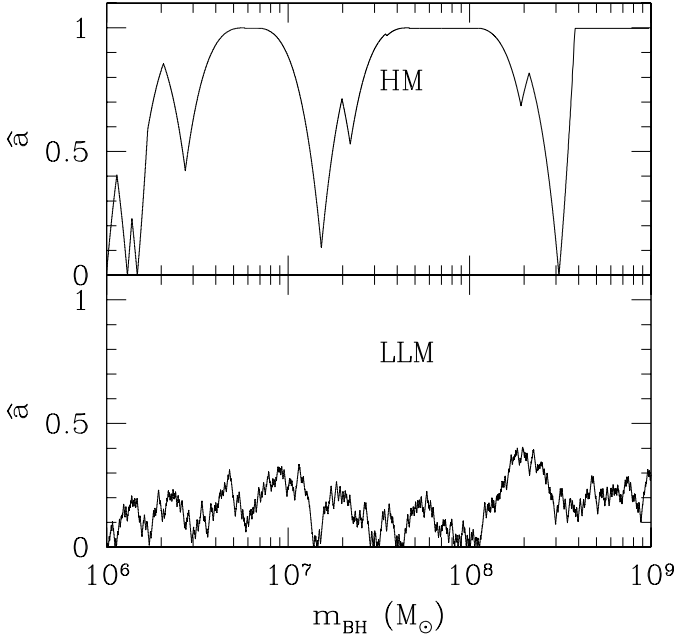}
\caption{Evolution of a BH spin during a series of accretion episodes lasting for a total of a Hubble time.
Initial mass $m_{\rm BH}=10^6 M_\odot$ , initial spin $\hat a=10^{-3}$. Lower panel:  the accreted mass at every accretion episode is constrained to be less than 0.01 of the BH mass (LLM). Upper panel: the accreted mass is randomly extracted in the range 0.01-10 times the BH mass (HM). Adapted from Volonteri, Sikora \& Lasota 2007.}
\label{fig3}
\end{figure}
 
Although observations of the iron line with the Chandra and XMM X-ray satellites are extending the studies of the innermost regions of BHs, aiming at probing black hole properties, interpretation of these studies is impeded by the inherent 'messiness' of gas dynamics. A far cleaner probe would be a compact mass in a precessing and decaying orbit around a massive hole. The detection of gravitational waves from a stellar mass BH, or even a white dwarf, or neutron star, falling into a massive black hole (Extreme Mass Ratio Inspiral, or EMRI)  can provide a unique tool to constrain the geometry of spacetime around BHs, and as a consequence, BH spins.  Detecting and monitoring such a system may have to await the launch of the planned {\it Laser Interferometer 
Space Antenna} ({\it LISA}).  Indeed the spin is a measurable parameter, with a very high accuracy, in the gravitational waves {\it LISA} signal (Barack \& Cutler 2004, Berti et al. 2004, Lang \& Hughes 2006, Vecchio 2004, Berti et al. 2006). Gravitational waves from an EMRI can be used to map the spacetime of the central massive dark object. The resulting 'map' can tell us if the standard picture for the central massive object, a Kerr BH described by general relativity, holds. 

 \section{Hierarchies of galaxies and black holes} 
The demography of massive dark objects, which we will continue to refer to as black holes in the following, has been clarified in the last ten years by studies of the central regions of relatively nearby galaxies (mainly with quiescent nuclei).  Strikingly, the centres of all observed galactic bulges host SMBHs (Richstone 2004).
 The mass of the quiescent SMBHs detected in the local Universe scales with the bulge luminosity - or stellar velocity dispersion - of their host galaxy (Ferrarese \& Merritt 2000; Gebhardt et al. 2000; Tremaine et al. 2002).  In the currently favoured cold dark matter cosmogonies (Spergel et al. 2006), present-day galaxies have been assembled via a series of mergers, from small-mass building blocks which form at early cosmic times.  In this paradigm galaxies experience multiple mergers during their lifetime. If most  galaxies host BHs in their centre, and a local galaxy has been made up by multiple mergers, then a black hole binary is a natural evolutionary stage.  After each merger event, the central black holes already present in each galaxy would be dragged to the centre of the newly formed galaxy  via dynamical friction, and then if/when they get close ($\approx 0.01-0.001$ pc) the black hole binary would coalesce  via emission of gravitational radiation.  
 
 The efficiency of dynamical friction decays when the BHs get close and form a binary, when the binary separation is around 0.1-1 pc (for $M_{BH}\simeq10^5-10^8 M_\odot$). Emission of gravitational waves becomes efficient at binary separations about two orders of magnitude smaller.  
In gas-poor systems, the subsequent evolution of the binary, while gravitational radiation emission is negligible,  may be largely determined by three-body interactions with background stars (Begelman, Blandford \& Rees 1980), by capturing the stars that pass within a distance of the order of the binary semi-major axis and ejecting them at much higher velocities (Quinlan 1996, Milosavljevic \& Merritt 2000, Sesana, Haardt \& Madau 2006). Dark matter particles will be ejected by decaying binaries in the same way  as the stars, i.e. through the gravitational slingshot. In minihalos a numerous population of low-mass stars may be present if the IMF were bimodal, with a second peak at 1-2 $M_\odot$, as suggested by Nakamura \& Umemura (2001). Otherwise the binary will be losing orbital energy to the dark matter background.  The hardening of the binary modifies the density profile, removing mass interior to the binary orbit, depleting the galaxy core of stars and dark matter, and slowing down further decay. 
In gas rich systems, however, the orbital evolution of the central SMBH is likely dominated by dynamical friction against the surrounding gaseous medium.  The available simulations (Mayer et al. 2006; Dotti et al. 2006a,b; Escala et al. 2004) show that the binary can shrink to about parsec or slightly subparsec scale by dynamical friction against the gas, depending on the gas thermodynamics. The interaction between a BH binary and an accretion disc can also lead to a very efficient transport of angular momentum, and drive the secondary BH to the regime where emission of gravitational radiation dominates on short timescales, comparable to the viscous timescale (Armitage \& Natarajan 2005, Gould \& Rix 2000).

 When the members of a black hole binary coalesce, there is a recoil due to the non-zero net linear momentum carried away by gravitational waves in the coalescence. If the holes have unequal masses, a preferred longitude in the orbital plane is determined by the orbital phase at which the final plunge occurs. For spinning holes there may be a rocket effect perpendicular to the orbital plane, since the spins break the mirror symmetry with respect to the orbital plane. This recoil could be so violent that the merged hole breaks loose from shallow potential wells, especially in the small mass pregalactic building blocks. 
 \begin{figure}
\includegraphics[height=3.5in,width=3.5in]{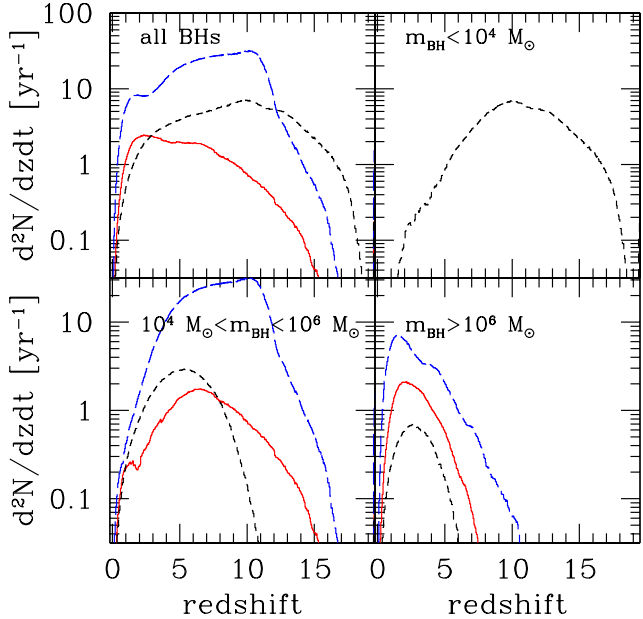}
  \caption{Predicted rate of BH binary coalescences per unit redshift, in different BH mass intervals. {\it Solid curve:} BH seeds from PopIII stars. {\it Long dashed curve:} BH seeds from direct collapse (Koushiappas et al. 2004).  {\it Short dashed curve:} BH seeds from direct collapse (Begelman, Volonteri \& Rees 2006). Adapted from Sesana et al. 2006.}
\end{figure}

 A single big galaxy can be traced back to the stage when it was in hundreds of smaller components with individual internal velocity dispersions as low as 20 ${\rm km \,s^{-1}}$. Did black holes form with the same efficiency in small galaxies (with shallow potential wells), or did their formation had to await the buildup of substantial galaxies with deeper potential wells? This issue is important because it determines the expected event rate detected by LISA, and whether there is a population of high-z miniquasars. Of particular interest is whether the merger history can be traced back to 'seed' holes, and be used to distinguish between seed formation scenarios. One first possibility is the direct formation of a BH from a collapsing gas cloud (Haehnelt \& Rees 1993;  Loeb \& Rasio 1994; Eisenstein \& Loeb 1995; Bromm \& Loeb 2003; Koushiappas, Bullock \& Dekel 2004; Begelman, Volonteri \& Rees 2006; Lodato \& Natarajan 2006). When the angular momentum of the gas is shed efficiently, on timescales shorter than star formation, the gas can collect in the very inner region of the galaxy. The loss of angular momentum can be driven either by (turbulent) viscosity or by global dynamical instabilities, such as the "bars-within-bars" mechanism (Shlosman, Frank \& Begelman 1989).  The infalling gas can therefore condense to form a central massive object.  The mass of the seeds predicted by different models vary, but typically are in the range $M_{BH} \sim 10^4-10^6\,M_\odot$. 

Alternatively, the seeds of SMBHs can be associated with the remnants of the first generation of stars, formed out of zero metallicity gas. The first stars are believed to form at $z\sim 20-30$ in halos which represent high-$\sigma$ peaks of the primordial density field.  The main coolant, in absence of metals, is molecular hydrogen, which is a rather inefficient coolant.  The inefficient cooling might lead to a very top-heavy initial stellar mass function, and in particular to the production of  an early generation of 'Very Massive Objects (VMOs)' (Carr, Bond, \& Arnett 1984). If very massive stars form above 260 $M_\odot$,  they would rapidly collapse to massive BHs with little mass loss (Fryer, Woosley, \& Heger 2001), i.e., leaving behind seed BHs with masses $M_{BH} \sim 10^2-10^3\,M_\odot$ (Madau \& Rees 2001; Volonteri, Haardt \& Madau 2003). 
 
LISA in principle is sensible to gravitational waves from binary BHs with masses in the range $10^3-10^6\; M_\odot$ basically at any redshift of interest.   A large fraction of coalescences will be directly observable by {\it LISA}, and on the basis of the detection rate, constraints can be put on the BH formation process. Different theoretical models for the formation of BH seeds and dynamical evolution of the binaries predict merger rates that largely vary one from the other (Figure 2).   {\it LISA} will be a unique probe of the formation and merger history of BHs along the {\it entire} cosmic history of galactic structures.

\section{Accretion} 
Accretion is inevitable during the 'active' phase of a galactic nucleus. Observations tell us that AGN are widespread in both the local and early Universe. All  the information that we have gathered on the evolution of SMBHs is indeed due to studies of AGN, as we have to await for LISA to be able to "observe" quiescent SMBHs in the distant Universe. A key issue is the relative importance of mergers and accretion in the build-up of the largest holes in giant ellipticals. 

The accretion of mass at the Eddington rate causes the black hole mass to increase in time as
\begin{equation}
M(t)=M(0)\,\exp\left(\frac{1-\epsilon}{\epsilon}\frac{t}{t_{\rm Edd}}\right),
\end{equation}
where $t_{\rm Edd}=0.45\,{\rm Gyr}$ and $\epsilon$ is the radiative efficiency. 
The classic argument of Soltan (1982), compares the total mass of black holes today with the total radiative output by known quasars,  by integration, over redshift and luminosity, of the luminosity function of quasars (Yu \& Tremaine 2002; Elvis, Risaliti, \& Zamorani 2002; Marconi et  al. 2004).  The total energy density can be converted  into the total mass density accreted by black holes during the active phase, by assuming a mass-to-energy conversion efficiency,  $\epsilon$ (Aller \& Richstone 2002; Merloni, Rudnick \& Di Matteo 2004).  The similarity of the total mass in SMBHs today and the total mass accreted by BHs  implies that the last 2-3 e-folds of the mass is grown via radiatively efficient accretion, rather than accumulated through mergers or radiatively inefficient accretion.  However,  most ot the Ôe-foldsÕ  (corresponding to a relatively small amount of mass, say the first  10\% of mass) could be gained rapidly via, e.g.,  radiatively inefficient accretion.  This argument is particularly important at early times.

\begin{figure}
\includegraphics[height=3.5in,width=3.5in]{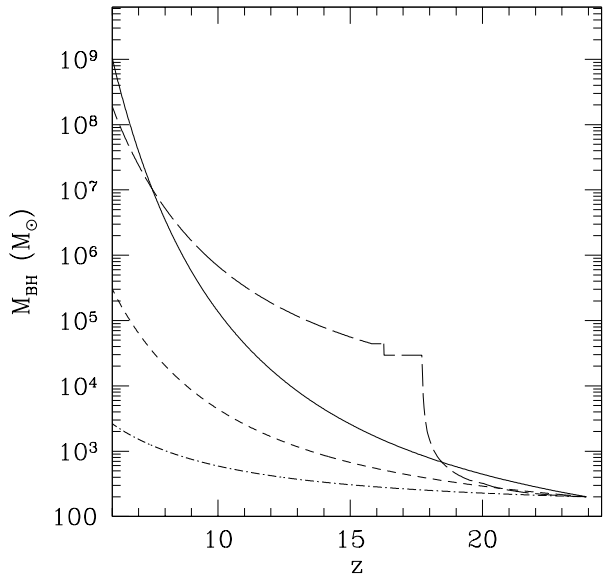}
  \caption{Growth of a BH mass under different assumption for the accretion
rate and efficiency. Eddington limited accretion: $\epsilon=0.1$ ({\it solid line}),
$\epsilon=0.2$ ({\it short dashed line}), $\epsilon=0.4$ ({\it dot-dashed line}). Radiatively inefficient super-critical accretion, as in Volonteri \& Rees 2005 ({\it long dashed line}). }
\end{figure}

The Sloan Digital  Sky survey detected luminous quasars at very high redshift, $z>6$, when the Universe was less than {\rm 1 Gyr} old. Follow-up observations confirmed that at least some of  these quasars are powered by supermassive black holes (SMBHs) with masses $\simeq 10^9\, M_\odot$ (Barth et al. 2003; Willott et al. 2003).  Given a seed mass $M(0)$ at $z=30$ or less, the higher the efficiency, the longer it takes for the BH to grow in mass by (say) 10 e-foldings. If accretion is radiatively efficient, via a geometrically thin disc, the alignment of a SMBH with the angular momentum of the accretion disc tends to efficiently spin holes up (see section 2) , and radiative efficiencies can therefore approach 30-40\%. With such a high efficiency, $\epsilon=0.3$, it can take longer than  {\rm 2 Gyr} for the seeds to grow up to a billion solar masses.   

Let us consider the extremely rare high redshift (say $z>15$) metal--free halos with virial temperatures $T_{\rm vir} > 10^4$K where gas can cool even in the absence of ${\rm H_2}$ via neutral hydrogen atomic lines. The baryons can therefore collapse until angular momentum becomes important. 
Afterward, gas settles into a rotationally supported dense disc at the center of the halo 
(Mo, Mao \& White 1998, Oh \& Haiman 2002). This gas can supply fuel for accretion onto a BH within it.  
Estimating the mass accreted by the BH within the Bondi-Hoyle formalism, the accretion rate is initially 
largely above the Eddington limit (Volonteri \& Rees 2005).  When the supply is super-critical the excess radiation can be trapped, as radiation pressure cannot prevent the accretion rate from being super-critical, while the emergent luminosity is still Eddington limited in case of spherical or quasi-spherical configurations (Begelman 1979; Begelman \& Meier 1982).  In the spherical case, 
Though this issue remains unclear, it still seems  possible that when the inflow rate is 
super-critical, the radiative efficiency drops so that the hole can accept the material without 
greatly exceeding the Eddington luminosity.  The  efficiency could be low either because most 
radiation is trapped and advected inward, or because the flow adjusts so that the material can 
plunge in from an orbit with small binding energy (Abramowicz \& Lasota 1980). 
The creation of a radiation-driven outflow, which can possibly stop the infall of material, is also a possibility.  If radiatively inefficient supercritical accretion requires metal-free conditions in exceedingly  rare massive halos, rapid early growth, therefore, can happen only for a tiny fraction 
of BH seeds. These SMBHs are those powering the $z=6$ quasars, and later on to be found in the most 
biased and rarest halos today. The global BH population, instead, evolves at a more quiet and slow pace.

\end{document}